# Superimposed Hi-C: A Solution Proposed for Identifying Single Cell's Chromosomal Interactions

Jia Zhang[#,1,2], Li Xiao[#,2], Peng Qi[1,3], Yaling Zeng[1,3], Xumeng Chen[1,3], Duan-fang Liao[*,1,4], Kai Li[*,1,2]

[#]These authors contributed equally to this work

[*]Correspondence: Kai Li, kaili34@yahoo.com ; Duan-fang Liao, dfliao@hnucm.edu.cn

1. National Engineering Research Center for Individualized Diagnosis & Treatment, Hunan University of Chinese Medicine, Changsha, 410208, Hunan, PR China
2. The Second Affiliated Hospital of Soochow University, Suzhou, 215004, PR China
3. Division of Stem Cell Regulation and Application, Key Laboratory for Quality Evaluation of Bulk Herbs of Hunan Province, Hunan University of Chinese Medicine, Changsha, Hunan 410208, China
4. Epigenetic laboratory, Changsha Health College, Liuyang economic development zone, Changsha, Hunan 410329, China

**Abstract:** Hi-C sequencing is widely used for analyzing chromosomal interactions. In this study, we propose "superimposed Hi-C," which features paired *EcoP15*I sites in a linker to facilitate sticky-end ligation with target DNAs. Superimposed Hi-C overcomes Hi-C's technical limitations, enabling the identification of single cell's chromosomal interactions.

**INTRODUCTION**

Chromosomal interactions play a crucial role in epigenetic mechanisms [1-3]. Currently, Hi-C sequencing is commonly employed to identify chromosomal interactions in samples consisting of tens of thousands of cells [4]. However, there is an urgent need for new technologies to analyze chromosomal interactions at the single-cell level, enhancing our understanding of temporal and spatial regulation of cellular processes [5-8].

As depicted in Figure 1, the library for superimposed Hi-C sequencing (Super Hi-C) can be efficiently constructed. Unlike conventional Hi-C, which involves multiple steps (such as enzymatic digestion, blunted-end preparation, and ligation between blunted ends), Super Hi-C employs sticky-end ligation. Specifically, a specialized linker containing a pair of EcoP15I sites is positioned tail-to-tail, facilitating ligation with crosstalked DNA targets [9-10]. This streamlined procedure significantly increases efficiency compared to conventional Hi-C.

The specialized linker used in Super Hi-C offers significant advantages, making it an ideal fit for identifying chromosomal interactions in single cell. The advantages are listed as follows: **i). Efficiency:** In conventional Hi-C, the ligation between crosstalked DNAs in a single cell must be highly efficient, as it cannot be compensated by the large number of cells typically used. Super Hi-C directly ligates crosstalked targets with the linker between their sticky termini. Sticky-end ligation is thousands of times more efficient than ligation between blunted ends. **ii). Flexibility:** During the ligation reaction in Hi-C, crosstalked termini are less flexible due to their partial association with denatured histones. As a result, ligation between these blunted ends tends to occur face-to-face spatially rather than side-by-side (as expected in Hi-C sequencing). The specialized linker eliminates this bias, favoring ligation with target termini in a side-by-side position. This ensures more space for annealing between sticky ends before ligation. **iii). Irreversible Ligation:** The ligation between the specialized linker and target termini is designed to be irreversible, unlike self-ligation between target termini, which is reversible. This design significantly improves specificity compared to conventional Hi-C. **iv). Increased Copy Number:** Super Hi-C uses a large copy number of linkers, enhancing the efficiency of ligation with crosstalked targets. And **v). Streamlined Procedure:** The use of the specialized linker eliminates several steps in conventional Hi-C, contributing to the overall increased efficiency of Super Hi-C. In summary, Super Hi-C's specialized linker addresses technical limitations and provides a powerful tool for studying chromosomal interactions at the single-cell level.

When there is a chromosomal interaction in which the DNA is recognizable by the endonuclease used, the paired *EcoP15*I sites residing in the linker can anchor as long as 27-nucleotide strings for each of the targets. Enzymatically digested products from *EcoP15*I have sticky ends with 5' extruded two nucleotides [8-9]. Compatible Illumina next-generation adapters are employed in constructing Super Hi-C sequencing libraries. Compared to constructing conventional Hi-C libraries with a low-efficient T/A ligation strategy, Super Hi-C sequencing libraries are prepared with a much more efficient ligation of two nucleotides extruded from sticky termini. As shown in Figure 1, the product from the ligation between targets and sequencing adapters is directly subjected to PCR reaction and sequencing, significantly shortening the library preparation procedures and thereby greatly increasing overall efficiency.

In summary, with the introduction of the specialized linker harboring a pair of *EcoP15*I sites, Super Hi-C substantially shortens the procedure for library construction from days to hours compared to conventional HI-C. It exponentially increases the ligation efficiency of crosstalked target DNA termini and overall efficiency in identifying chromosomal interactions (Table1). For the analysis of chromosomal interactions in samples consisting of a large number of cells, the application of Super Hi-C sequencing provides a time-efficient, less costly option with exponentially increased efficiency. These technical advantages over conventional HI-C position this new method as a solution for identifying single-cell chromosomal interactions. Super Hi-C is expected to be a powerful tool in our efforts to establish the full spectrum of chromosomal interactions in human cells at the single-cell level.

**Table 1. Comparison of Library Construction Procedures Between Conventional and Superimposed Hi-C Sequencing**

| Hi-C | Digestion-ligation-only Hi-C | Super HI-C | Compare |
|---|---|---|---|
| 1.Crosslink DNA 2.Cut with restriction enzyme 3.Fill ends and mark with biotin 4.Ligate 5.Purify and sheared DNA；pull down biotin 6.Sequence with paired-ends | 1.Nucleus double cross-linking 2.Chromatin digestion 3.Simultaneous digestion and ligation of half-linkers 4.In-gel proximity ligation 5.MmeI digestion 6.High-throughput sequencing | 1.Crosslink DNA 2.Digestion and linker ligation 3.EcoP15I digestion 4.Ligation NGS adapter and sequence | 1.Compared with Hi-C and DLO Hi-C, Super Hi-C is easy to operate, shortens operation time, and saves costs 2.Compared with Hi-C, remove biotin labeling; Changing the flat end connection to a sticky end linker improves detection efficiency 3. Compared with DLO Hi-C, the semi linker connection is changed to a fully linker connection, and the one-time connection is directly introduced into *EcoP15*I; Replace *Mme*I with |

| | | *Ecop15*I, which has a longer recognition site; After enzyme digestion, directly connect to NGS adapter without gel recovery step |

In this table, we compare the steps involved in library construction for conventional Hi-C and superimposed Hi-C sequencing. The efficiency-restrained steps in conventional Hi-C are briefly summarized, along with their potential causes. Notably, blunted end ligation, which is crucial for identifying chromosomal interaction sites, represents the least efficient step. To compensate for this, researchers often increase cell numbers to the level of thousands. The overall low efficiency of conventional Hi-C arises from several factors, including DNA extraction, DNA fragmentation, biotin enrichment of target DNA, and T/A ligation to integrate next-generation adapters. In contrast, superimposed Hi-C eliminates all efficiency-restrained steps, making it highly sensitive for analyzing chromosomal interactions in single cells.

# Figure 1. Superimposed Hi-C Method

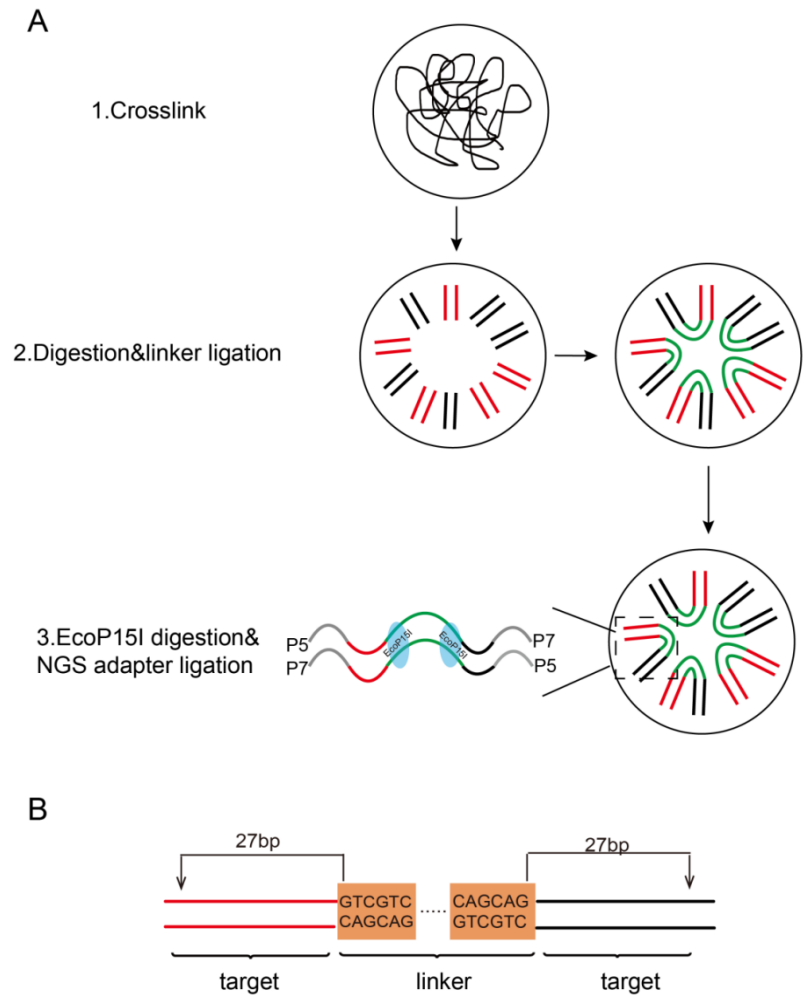

# Legends

## Figure 1. Superimposed Hi-C Method

**Panel A:** This illustration depicts the shortened procedure for constructing a Hi-C sequencing library using the new superimposed Hi-C method. Compared to conventional Hi-C, the superimposed Hi-C library can be completed within hours instead of days. This efficiency is largely attributed to the introduction of a specialized linker that ligates the termini of DNA fragments at chromosomal interaction sites after digestion by selected endonuclease(s).

**Panel B:** Here, we visualize the possible combinations of target DNA termini and the linker. The design of the specialized linker not only eliminates discrimination but also promotes ligation between termini positioned side by side in conventional Hi-C.